# Bose-Einstein condensation in perfect crystals

V. A. Golovko

Moscow State Evening Metallurgical Institute

Lefortovsky Val 26, Moscow 111250, Russia

E-mail: mgvmi-mail@mtu-net.ru

Abstract

To investigate the phenomenon of Bose-Einstein condensation in perfect crystals a hierarchy of equations for reduced density matrices that describes a thermodynamically equilibrium quantum system is employed, the hierarchy being obtained earlier by the author. The thermodynamics of a crystal with a condensate and the one of a crystal with no condensate are constructed in parallel, which is required for studying the phase transition involving Bose-Einstein condensation. The transition is analysed also with the help of the Landau theory of phase transitions which shows that a superfluid state can result either from two consecutive phase transitions or from only one. To demonstrate how the general equations obtained can be applied for a concrete crystal the bifurcation method for solving the equations is utilized. New results concerning properties of the condensate crystals at zero temperature are obtained as well. In the concluding section, the physical concept of the condensate is discussed.

**1. Introduction**

In previous papers [1,2] (hereafter referred to as I and II respectively), perfect crystals below the Bose-Einstein condensation (BEC) point that can be superfluid were investigated with the help of the approach in equilibrium quantum statistical mechanics proposed in Ref. [3] (see also [4]). The approach is based upon a hierarchy of equations for reduced density matrices obtained from the quantum mechanical Liouville equation. However, the studies of I and II were restricted mainly to the case of absolute zero of temperature where the condensate comprises all particles while the phenomenon *per se* of BEC in the crystals was not considered in those papers. The aim in the present paper is to investigate the phenomenon, for which thermodynamics is required to treat nonzero temperatures. It should be emphasized that BEC in a perfect crystal is implied in the present paper whereas it is common practice to connect BEC in a solid with disruptions of the ideal crystalline order (vacancies, interstitials) (see, e.g., Ref. [5]).

A characteristic feature of the aforementioned approach is construction of thermodynamics compatible with the hierarchy without use made of the standard Gibbs method [3,4]. The relevant thermodynamics of an ordinary (with no condensate) quantum crystal was constructed in Ref. [6]. In that reference, however, in order to simplify the matter and to avoid solving a rather involved equation for a function $\tau(\theta,\rho_0)$ (here and henceforth $\theta$ is the temperature in units of energy, $\theta = k_B T$, and $\rho_0$ is the average number density of particles), a simplified form for another important function $n(z)$ that contains $\tau(\theta,\rho_0)$ (see below) was employed, a form without $\tau(\theta,\rho_0)$. Although the approximation exploited in [6] may describe properties of a quantum crystal sufficiently well, it is unsuitable for our purposes because the function $\tau(\theta,\rho_0)$ plays an important role in the matter of BEC [7]. For this reason, in Sec. 2 of this paper we first work out thermodynamics of an ordinary quantum crystal with use made of the exact form of $n(z)$, which is necessary as a preliminary step for treatment of crystals with a condensate (condensate crystals).

In Sec. 3 we discuss two different order parameters that figure in the case of a superfluid condensate crystal and analyse the situation with the help of the Landau theory of phase transitions. Section 4 is devoted to a crystal below the BEC point. We first write down and investigate relevant equations and then elaborate thermodynamics.

In order to demonstrate how the general equations obtained can be applied for a concrete crystal, in Sec. 5 we consider the bifurcation method for solving the equations describing the crystal restricting ourselves to the example of a simple cubic lattice, in which case the equation for the condensate can be reduced to the well-known Mathieu equation (in a certain approximation).

In Sec. 6 we consider a mathematical method that enables one to shed a new light on properties of the condensate crystals at zero temperature in addition to the ones obtained in II.

For the sake of convenience, when referring to an equation of papers I or II we shall place I or II in front; so we shall write, e. g., (I.41) implying equation (41) of I or (II.10) implying equation (10) of II.

## 2. Ordinary quantum crystal

As in I and II we consider a system of $N$ spinless bosons enclosed in a volume $V$. The particles of mass $m$ interact via a two-body potential $K(|\mathbf{r}_j - \mathbf{r}_k|)$. Let us write down the first equations of the equilibrium quantum hierarchy obtained in Ref. [3] (the equations are presented in [6] with comments):

$$\rho(\mathbf{r}_1)\,\nabla_1 U(\mathbf{r}_1) = \int_V \rho_2(\mathbf{r}_1,\mathbf{r}_2)\,\nabla_1 K(|\mathbf{r}_1 - \mathbf{r}_2|)\,d\mathbf{r}_2\,, \tag{1}$$

$$\rho(\mathbf{r}) = \frac{1}{(2\pi)^4\,i\hbar^3}\int_{(\infty)} d\mathbf{p} \int_C dz\, n(z)\, v(\mathbf{r},\mathbf{p},z), \tag{2}$$

$$\frac{\hbar^2}{2m}\nabla^2 v(\mathbf{r},\mathbf{p},z) + \frac{i\hbar}{m}\mathbf{p}\nabla v + \left[z - \frac{\mathbf{p}^2}{2m} - U(\mathbf{r})\right] v = 1. \tag{3}$$

Here $\rho(\mathbf{r})$ and $\rho_2(\mathbf{r}_1,\mathbf{r}_2)$ are diagonal elements of the singlet and pair density matrices respectively, $\rho(\mathbf{r})$ yielding the spatial number density of particles periodic in the event of a crystal. By (2) the density $\rho(\mathbf{r})$ depends upon an effective potential $U(\mathbf{r})$ via an auxiliary function $v(\mathbf{r},\mathbf{p},z)$ satisfying Eq. (3) while the potential $U(\mathbf{r})$ itself is determined by Eq. (1). The integration in (2) over the complex variable $z$ is carried out along a contour $C$ that should enclose all poles of $v(\mathbf{r},\mathbf{p},z)$ as a function of $z$ (the contour is shown in Fig. 1 of [3]). The density $\rho(\mathbf{r})$ is normalized according to

$$\int_V \rho(\mathbf{r})\,d\mathbf{r} = N. \tag{4}$$

For use later we write down also the singlet density matrix:

$$R_1(\mathbf{r},\mathbf{r}') = \frac{1}{(2\pi)^4\,i\hbar^3}\int_{(\infty)} d\mathbf{p} \int_C dz\, n(z)\, v(\mathbf{r},\mathbf{p},z)\exp\left[\frac{i}{\hbar}\mathbf{p}(\mathbf{r}-\mathbf{r}')\right]. \tag{5}$$

Note that $\rho(\mathbf{r}) = R_1(\mathbf{r},\mathbf{r})$ in accord with (2). The function $n(z)$ in (2) and (5) is [3]

$$n(z) = B e^{-z/\tau}, \tag{6}$$

where we write $B$ instead of $A$, the constant $B$ being fixed by the normalization of (4). The quantity $\tau$ in (6) is a function of the temperature $\theta$ and the average density $\rho_0 = N/V$ (see also Introduction). An equation for $\tau(\theta,\rho_0)$ should follow from thermodynamics.

The diagonal element of the pair density matrix that figures in (1) can be represented as

$$\rho_2(\mathbf{r}_1,\mathbf{r}_2) = \rho(\mathbf{r}_1)\,\rho(\mathbf{r}_2)\,g(\mathbf{r}_1,\mathbf{r}_2), \tag{7}$$

where $g(\mathbf{r}_1,\mathbf{r}_2)$ is a pair correlation function. The function $\rho_2(\mathbf{r}_1,\mathbf{r}_2)$ and consequently $g(\mathbf{r}_1,\mathbf{r}_2)$ can be found from further equations of the hierarchy of [3], for which the hierarchy should be closed somehow. This is a well-known problem characteristic of the classical BBGKY hierarchy (recall that Eq. (1) and further equations for the effective potentials are fully analogous with the ones that follow from the BBGKY hierarchy while the quantum hierarchy itself goes over into the classical BBGKY hierarchy in the limit as $\hbar \to 0$ [3,4]). The form of the pair correlation function $g(\mathbf{r}_1,\mathbf{r}_2)$ for crystals is rather involved [8]. At the same time, the function $g(\mathbf{r}_1,\mathbf{r}_2)$ plays a secondary role in the crystals because then the structure of $\rho_2(\mathbf{r}_1,\mathbf{r}_2)$ as given by (7) is determined, first of all, by the peak-like periodic density $\rho(\mathbf{r})$, as distinct from fluids where $\rho(\mathbf{r})$ = constant and thereby $g(\mathbf{r}_1,\mathbf{r}_2)$ is the governing function in $\rho_2(\mathbf{r}_1,\mathbf{r}_2)$ [8,9]. Practically in all studies on statistical theory of crystals, for the sake of simplicity one assumes that the pair correlation function $g(\mathbf{r}_1,\mathbf{r}_2)$ depends only on $|\mathbf{r}_1 - \mathbf{r}_2|$:

$$g(\mathbf{r}_1,\mathbf{r}_2) = g(|\mathbf{r}_1 - \mathbf{r}_2|,\,\theta,\,\rho_0). \tag{8}$$

It may be remarked that in Ref. [6] a special form of (8) was utilized. In what follows we shall presume the function $g(\mathbf{r}_1,\mathbf{r}_2)$ to be given by (8) without further specification.

The form of (8) is convenient in that Eq. (1) is readily integrated to yield

$$U(\mathbf{r}) = \int_V K_g(|\mathbf{r} - \mathbf{r}'|)\,\rho(\mathbf{r}')\,d\mathbf{r}', \tag{9}$$

in which (for brevity, the parameters $\theta$ and $\rho_0$ are omitted in $g(r) \equiv g(|\mathbf{r}|)$)

$$K_g(r) = \int_\infty^r \frac{dK(r')}{dr'}\,g(r')\,dr'. \tag{10}$$

Implying the crystalline state we look for periodic solutions of the above equations in terms of Fourier series

$$\rho(\mathbf{r}) = \sum_{l,m,n=-\infty}^{\infty} a_{lmn}\,e^{i\mathbf{A}\mathbf{r}}, \qquad v(\mathbf{r},\mathbf{p},z) = \sum_{l,m,n} b_{lmn}(\mathbf{p},z)\,e^{i\mathbf{A}\mathbf{r}}, \tag{11}$$

where $\mathbf{A} = l\mathbf{a}_1 + m\mathbf{a}_2 + n\mathbf{a}_3$ with the basic reciprocal-lattice vectors $\mathbf{a}_1$, $\mathbf{a}_2$ and $\mathbf{a}_3$ (for detail see Ref. [10]). Substituting this into (9) gives

$$U(\mathbf{r}) = \sum_{l,m,n} a_{lmn}\,\sigma(A)\,e^{i\mathbf{A}\mathbf{r}} \tag{12}$$

with

$$\sigma(k) = \int K_g(|\mathbf{r}|) e^{i\mathbf{k}\mathbf{r}} d\mathbf{r} = \frac{4\pi}{k} \int_0^\infty r K_g(r) \sin kr \, dr, \qquad (13)$$

where $k = |\mathbf{k}|$ (we imply that $V \to \infty$).

Upon substituting all these expansions into (2) and (3) we obtain the following set of equations for $a_{lmn}$ and $b_{lmn}(\mathbf{p},z)$

$$\left[ z - \frac{(\mathbf{p} + \hbar \mathbf{A})^2}{2m} \right] b_{lmn} - \sum_{l',m',n'} a_{l'm'n'} \, \sigma(A') b_{l-l' \, m-m' \, n-n'} = \delta_{lmn}^{000}, \qquad (14)$$

$$a_{lmn} = \frac{B}{(2\pi)^4 i\hbar^3} \int_{(\infty)} d\mathbf{p} \int_C b_{lmn}(\mathbf{p},z) e^{-z/\tau} dz, \qquad (15)$$

where $\delta_{lmn}^{000}$ is the three-dimensional Kronecker symbol. The constant $B$ is specified by the average number density $\rho_0 = N/V$ on account of the fact that $\rho_0 = a_{000}$. In the following we shall write $a_0$ for $a_{000}$ and $b_0$ for $b_{000}$. It is worth remarking that Eq. (14) can be solved by iteration yielding equation (3.9) of Ref. [6]. In Appendix A we shall consider another method for treating Eq. (3).

We turn now to construction of thermodynamics compatible with the above hierarchy equations in line with the ideas of Ref. [3] (see also [4]). At this stage of the theory, we are in a position to find expressions for the internal energy $E$ and the stress tensor $\sigma_{ij}$. The expressions for $E$ and $\sigma_{ij}$ must obey the laws of thermodynamics for equilibrium (quasi-static) processes. This requirement leads to an equation for the function $\tau(\theta,\rho_0)$ that figures in (6) because $E$ and $\sigma_{ij}$ depend upon $\tau(\theta,\rho_0)$. In case the equation is satisfied, one will be able to calculate wanted thermodynamic functions and quantities by leaning on standard thermodynamic formulae.

The internal energy $E$ of a crystal is written down in equation (4.1) of [6] without derivation. The first term is obtained when (5) is inserted into (I.41) and the result is averaged over the volume of the crystal with use made of (11). The second term is the same as in the classical case [8]. As a consequence,

$$E = V \int_{(\infty)} \frac{\mathbf{p}^2}{2m} \varphi(\mathbf{p}) \, d\mathbf{p} + \frac{V}{2} \sum_{l,m,n} |a_{lmn}|^2 \sigma_e(A), \qquad (16)$$

where

$$\varphi(\mathbf{p}) = \frac{B}{(2\pi)^4 i\hbar^3} \int_C b_0(\mathbf{p},z) e^{-z/\tau} dz, \qquad (17)$$

and

$$\sigma_e(k) = \int K(|\mathbf{r}|) \, g(|\mathbf{r}|) e^{i\mathbf{k}\mathbf{r}} d\mathbf{r} = \frac{4\pi}{k} \int_0^\infty r K(r) \, g(r) \sin kr \, dr. \qquad (18)$$

The quantum stress tensor $\sigma_{ij}$ is given by Eq. (II.10). The first term can be computed with use made of (5) again while the second term is the same as in the classical case [8]:

$$\sigma_{ij} = -\frac{1}{m}\int_{(\infty)} p_i p_j \varphi(\mathbf{p}) d\mathbf{p} - \frac{1}{2}\sum_{l,m,n}|a_{lmn}|^2\left\{\delta_{ij}\left[\sigma(A,\theta,\rho_0) - \rho_0\frac{\partial\sigma}{\partial\rho_0}\right] + a_{ki}\frac{\partial\sigma}{\partial a_{kj}}\right\}. \quad (19)$$

Here and henceforth, $a_{kl}$ are components of the basic reciprocal-lattice vectors $\mathbf{a}_1$, $\mathbf{a}_2$, $\mathbf{a}_3$ defined in [10], and summation from 1 to 3 over the repeated indices is implied. It should be remarked that the second term of (19) differs in appearance from the one of equation (4.6) of [6] (the last expression was used in (II.12) as well). The two expressions are identical if account is taken of the fact that $\sigma = \sigma(k,\theta,\rho_0)$ by (8) and

$$a_{ki}\frac{\partial\sigma(A,\theta,\rho_0)}{\partial a_{kj}} = \frac{A_i A_j}{A}\frac{\partial\sigma}{\partial A} + \delta_{ij}\rho_0\frac{\partial\sigma}{\partial\rho_0}. \quad (20)$$

It will be noted that equations (3.4) and (3.5) of [10] are implied in the present calculations.

With $E$ and $\sigma_{ij}$ at our disposal we turn to the laws of equilibrium thermodynamics. The first law stipulates that the quantity of heat $\delta Q$ received by a system is equal to $dE$ plus the work done by the system. The work for the crystal is specified by equation (4.15) of [10], so that

$$\delta Q = dE + \frac{V}{2\pi}\sigma_{ij} d_{ki} da_{kj}, \quad (21)]$$

where $d_{ki}$ are quantities defined in [10]. From the second law of thermodynamics it follows that $\delta Q = \theta dS$ where $S$ is the entropy. The function $S$ exists only if the compatibility conditions given by (4.19) of Ref. [10] hold. Similarly to (5.19) of Ref. [3], the first condition yields

$$\frac{a_{ki}}{V}\frac{\partial E}{\partial a_{kj}} + \sigma_{ij} - \theta\frac{\partial\sigma_{ij}}{\partial\theta} = 0. \quad (22)$$

The second condition acquires the form

$$\frac{\partial\sigma_{nj}}{\partial a_{kl}}d_{in} - (d_{kl}d_{in} + d_{il}d_{kn})\frac{\sigma_{nj}}{2\pi} = \frac{\partial\sigma_{nl}}{\partial a_{ij}}d_{kn} - (d_{ij}d_{kn} + d_{kj}d_{in})\frac{\sigma_{nl}}{2\pi} \quad (23)$$

and imposes certain restrictions on the pair correlation function $g(r,\theta,\rho_0)$ of (8). In the case of a cubic symmetry where $\sigma_{ij} = -p\delta_{ij}$ ($p$ is the pressure), this last condition is satisfied identically.

In what follows we shall imply solely cubic crystals. Then the condition of (22) remains alone. By making use of (16), (19) and (20) the condition can be cast as

$$\frac{\theta}{3m}\int\mathbf{p}^2\frac{\partial\varphi}{\partial\theta}d\mathbf{p} + \frac{\rho_0}{2m}\int\mathbf{p}^2\frac{\partial\varphi}{\partial\rho_0}d\mathbf{p} - \frac{5}{6m}\int\mathbf{p}^2\varphi(\mathbf{p})d\mathbf{p}$$

$$-\frac{1}{2}\sum_{l,m,n}|a_{lmn}|^2\left(\sigma_e(A) + \sigma(A) - \rho_0\frac{\partial\sigma_e}{\partial\rho_0} + \frac{A}{3}\frac{\partial\sigma}{\partial A} - \theta\frac{\partial\sigma}{\partial\theta} - \frac{\theta A}{3}\frac{\partial^2\sigma}{\partial\theta\partial A}\right)$$

$$+ \frac{\rho_0}{2} \sum_{l,m,n} \frac{\partial |a_{lmn}|^2}{\partial \rho_0} \sigma_e(A) + \frac{\theta}{2} \sum_{l,m,n} \frac{\partial |a_{lmn}|^2}{\partial \theta} \left( \sigma(A) + \frac{A}{3} \frac{\partial \sigma}{\partial A} \right) = 0. \qquad (24)$$

This equation is in fact a partial differential equation for $\tau(\theta,\rho_0)$ because the function $\varphi(\mathbf{p})$ depends on $\tau(\theta,\rho_0)$ via (17) (cf. equation (5.20) of [3]). The initial condition for the equation follows from the fact that, on the line (in the $\rho$–$\theta$ phase plane) of the phase transition from the fluid, the function $\tau(\theta,\rho_0)$ should coincide with the function $\tau(\theta,\rho)$ of the fluid if $\rho = \rho_0$. Therefore, to solve Eq. (24) it is necessary first to solve the corresponding equation for the fluid, namely, equation (5.20) of [3].

Once Eq. (24) is solved, one can calculate any thermodynamic function or quantity with use made of standard thermodynamic equations. For example, if the Helmholtz free energy $F = E - \theta S$ is required, it can be calculated with the help of the equations

$$\theta^2 \left( \frac{\partial}{\partial \theta} \frac{F}{\theta} \right)_{\rho_0} = -E, \qquad \left( \frac{\partial F}{\partial V} \right)_\theta = -\frac{\rho_0^2}{N} \left( \frac{\partial F}{\partial \rho_0} \right)_\theta = p. \qquad (25)$$

Herein the energy $E$ is given by (16) while the pressure $p$ can be found from (19) as long as $\sigma_{ij} = -p\delta_{ij}$ for the cubic symmetry. It is important to emphasize that these two simultaneous equations for $F$ are compatible by virtue of the fulfilment of the compatibility conditions for the entropy $S$. Different forms for $F$ obtainable on the basis of (25) can be found in Appendix B of Ref. [11]. In Sec. 4 below we shall deduce and solve another equation for $F$.

## 3. Landau-theory-based analysis

Before proceeding to condensate crystals let us discuss the problem under study from the viewpoint of order parameters. From the results of I it follows that in the present problem there are two different order parameters, namely, the condensate average density $\rho_c$ and the vector $\mathbf{p}_0$ that determines the macroscopic superflow $\mathbf{P}$ by (I.27). Therefore, there should exist two different phase transitions: first BEC where $\rho_c$ acquires a spontaneous value and subsequently, on lowering the temperature further, another phase transition to a superfluid state where $\mathbf{P}$ acquires a spontaneous value (the second phase transition may not happen down to zero temperature). The crystal symmetry should not change in the first phase transition while it does change in the second (see I).

At the same time, these order parameters are not fully independent of one another inasmuch as a state with $\rho_c = 0$ and $\mathbf{p}_0 \neq 0$ cannot be superfluid because from (I.27) it follows that, if $\rho_c = 0$, always $\mathbf{P} = 0$ regardless of $\mathbf{p}_0$. When the temperature is slightly above the BEC point and the free energy with $\rho_c \neq 0$ is nonequilibrium, this free energy can have a minimum at $\mathbf{p}_0 \neq 0$.

Consequently, when BEC occurs, the minimum of the free energy will correspond to $\mathbf{p}_0 \neq 0$ and the relevant condensate phase will be simultaneously superfliud. Just this last situation takes place in liquid helium which becomes superfluid precisely at the λ-transition.

The above reasoning can be confirmed by the Landau theory of phase transitions. Let us write down the Landau free energy [12]

$$F = F_0 + \alpha \eta^2 + \beta \eta^4, \tag{26}$$

where $F_0$ is a term irrelevant to our problem, η is the order parameter and $\beta > 0$ (in actual fact, $\eta = \pm\sqrt{\rho_c}$; see the full version of Ref. [7]). Thermodynamic quantities depend upon the vector $\mathbf{p}_0$ (see I, II and the next section) and thereby the coefficients α and β should depend upon $\mathbf{p}_0$ as well. The $\mathbf{p}_0$-dependence of β plays an insignificant role as long as we presume that always $\beta > 0$. The $\mathbf{p}_0$-dependence of α can be represented as

$$\alpha = \alpha_0 + \alpha_1 p_0^2 + \alpha_2 p_0^4. \tag{27}$$

For the sake of simplicity we neglect the anisotropy in case a crystal is concerned and assume that $\alpha_2 > 0$ as β in (26) and that only $\alpha_0$ depends on the temperature as is customary.

If $\alpha_1 > 0$, the minimum of $F$ corresponds to $p_0 = 0$, and at $\alpha = \alpha_0 = 0$ there occurs a phase transition (BEC) without superfluidity. Only afterwards there may happen another phase transition where $p_0$ takes a nonzero value.

If $\alpha_1 < 0$, the minimization of $F$ with α of (27) yields

$$p_0^2 = -\frac{\alpha_1}{2\alpha_2}. \tag{28}$$

When this value is inserted into (27) and (26), one obtains

$$F = F_0 + \tilde{\alpha}\eta^2 + \beta\eta^4 \quad \text{with} \quad \tilde{\alpha} = \alpha_0 - \frac{\alpha_1^2}{4\alpha_2}. \tag{29}$$

Now BEC occurs at $\tilde{\alpha} = 0$ whereas the condensate phase turns out superfluid with $p_0$ given by (28). At the same time the macroscopic superflow $\mathbf{P}$ will be proportional to $\rho_c$ according to (I.27), being nil at $\tilde{\alpha} = 0$.

It should be remarked that, when writing (27), we assumed that $p_0$ was small only for simplicity's sake. In the general case, BEC without superfluidity occurs given the minimum of $\alpha(p_0)$ in (26) at $p_0 = 0$, whereas the arising condensate phase will be superfluid if the minimum of $\alpha(p_0)$ is at $p_0 \neq 0$. The present paper is devoted mainly the first case where BEC occurs without attendant superfluidity.

## 4. Condensate crystal

In the instance of a condensate crystal, the reasoning that leads to (8) remains valid and we can use Eq. (8) as it stands. Therefore, from Eq. (1) that does not depend upon concrete forms of the functions that figure in it, one obtains (9) with (10). Now, however, the crystal density is, in view of (I.9),

$$\rho(\mathbf{r}) = \rho^{(c)}(\mathbf{r}) + \rho^{(n)}(\mathbf{r}). \tag{30}$$

The normal part of the density is described by an equation analogous to (2), namely,

$$\rho^{(n)}(\mathbf{r}) = \frac{B}{(2\pi)^4 i\hbar^3} \int_{(\infty)} d\mathbf{p} \int_C dz\, e^{-z/\tau}\, v(\mathbf{r},\mathbf{p},z), \tag{31}$$

while Eq. (3) for the new function $v(\mathbf{r},\mathbf{p},z)$ preserves its form.

The condensate part of the density $\rho^{(c)}(\mathbf{r})$ is specified by Eqs. (I.14) and (I.15) which, like (I.28) and (I.29), can be recast in the form

$$\rho^{(c)}(\mathbf{r}) = \rho_c |u(\mathbf{r})|^2 \tag{32}$$

with the following equation for $u(\mathbf{r})$

$$\frac{\hbar^2}{2m}\nabla^2 u(\mathbf{r}) + \frac{i\hbar}{m}\mathbf{p}_0 \nabla u(\mathbf{r}) + \left[\varepsilon_{(1)} - \frac{\mathbf{p}_0^2}{2m} - U(\mathbf{r})\right] u(\mathbf{r}) = 0. \tag{33}$$

The above equations of this section together with (3) and (9) constitute the set of equations that describes the condensate crystal. To this must be added the normalization conditions

$$\rho_c = \frac{1}{v_0}\int_{v_0} \rho^{(c)}(\mathbf{r}) d\mathbf{r}, \quad \rho_n = \frac{1}{v_0}\int_{v_0} \rho^{(n)}(\mathbf{r}) d\mathbf{r}, \quad \frac{1}{v_0}\int_{v_0}|u(\mathbf{r})|^2 d\mathbf{r} = 1, \tag{34}$$

where the integration is carried out over the volume of the elementary crystal cell $v_0$. It will be noted that $\rho_0 = \rho_c + \rho_n$ in accord with (30).

The densities $\rho^{(c)}(\mathbf{r})$ and $\rho^{(n)}(\mathbf{r})$ can be expanded in Fourier series similarly to (11) with respective coefficients $a_{lmn}^{(c)}$ and $a_{lmn}^{(n)}$. The expansion for $U(\mathbf{r})$ of (12) remains valid with

$$a_{lmn} = a_{lmn}^{(c)} + a_{lmn}^{(n)}. \tag{35}$$

It is convenient to expand also the function $u(\mathbf{r})$ of (32):

$$u(\mathbf{r}) = \sum_{l,m,n=-\infty}^{\infty} c_{lmn}\, e^{i\mathbf{A}\mathbf{r}} \quad \text{with} \quad a_{lmn}^{(c)} = \rho_c \sum_{l',m',n'} c_{l'm'n'} c^{*}_{l'-l,m'-m,n'-n}. \tag{36}$$

The internal energy of the crystal can be calculated with use made of (I.41) and (I.25). The energy due to the condensate can be transformed with the help of (36) while the one of the normal fraction is given by (16)–(18). As a result,

$$E = \frac{\rho_c V}{2m} \sum_{l,m,n} (\mathbf{p}_0 + \hbar \mathbf{A})^2 |c_{lmn}|^2 + V \int_{(\infty)} \frac{\mathbf{p}^2}{2m} \varphi(\mathbf{p}) d\mathbf{p} + \frac{V}{2} \sum_{l,m,n} |a_{lmn}|^2 \sigma_e(A). \qquad (37)$$

It should be remarked that the condensate energy is written here in a form other than in (I.42). The stress tensor can be computed by analogy with (II.12) and (19):

$$\sigma_{ij} = -\frac{\rho_c}{m} \sum_{l,m,n} (p_{0i} + \hbar A_i)(p_{0j} + \hbar A_j)|c_{lmn}|^2 - \frac{1}{m} \int_{(\infty)} p_i p_j \varphi(\mathbf{p}) d\mathbf{p}$$

$$- \frac{1}{2} \sum_{l,m,n} |a_{lmn}|^2 \left\{ \delta_{ij} \left[ \sigma(A,\theta,\rho_0) - \rho_0 \frac{\partial \sigma}{\partial \rho_0} \right] + a_{ki} \frac{\partial \sigma}{\partial a_{kj}} \right\}. \qquad (38)$$

In what follows we shall consider the condensate phase without superfluidity where $\mathbf{p}_0 = 0$ (see the end of the preceding section). In this case Eq. (37) reduces to

$$E = \frac{\rho_c V \hbar^2}{2m} \sum_{l,m,n} A^2 |c_{lmn}|^2 + V \int_{(\infty)} \frac{\mathbf{p}^2}{2m} \varphi(\mathbf{p}) d\mathbf{p} + \frac{V}{2} \sum_{l,m,n} |a_{lmn}|^2 \sigma_e(A). \qquad (39)$$

Implying cubic crystals where (23) holds automatically, we can employ the pressure $p$ instead of $\sigma_{ij}$ since then $\sigma_{ij} = -p\delta_{ij}$. From (38) with the help of (20) we get

$$p = \frac{\rho_c \hbar^2}{3m} \sum_{l,m,n} A^2 |c_{lmn}|^2 + \int_{(\infty)} \frac{\mathbf{p}^2}{3m} \varphi(\mathbf{p}) d\mathbf{p} + \frac{1}{2} \sum_{l,m,n} |a_{lmn}|^2 \left[ \sigma(A) + \frac{A}{3} \frac{\partial \sigma}{\partial A} \right]. \qquad (40)$$

It is interesting to observe that in the event of a fluid the condensate contributes neither to the kinetic part of the energy $E$ nor to the one of the pressure $p$ if $\mathbf{p}_0 = 0$, which is seen from equations (3.3), (3.4) and (3.7) of Ref. [7]. In the crystal, the kinetic contributions to $E$ and $p$ do exist even if $\mathbf{p}_0 = 0$ and are given by the first terms in (39) and (40) (the terms disappear for the fluid where only $c_{000} \neq 0$ to which $A = 0$ corresponds).

In the cubic case, Eq. (22) reduces to (cf. equation (5.19) of [3])

$$\frac{\rho_0^2}{N} \frac{\partial E}{\partial \rho_0} = p - \theta \frac{\partial p}{\partial \theta}, \qquad (41)$$

and yields the following equation for $\tau(\theta,\rho_0)$

$$\frac{\theta}{3m} \int \mathbf{p}^2 \frac{\partial \varphi}{\partial \theta} d\mathbf{p} + \frac{\rho_0}{2m} \int \mathbf{p}^2 \frac{\partial \varphi}{\partial \rho_0} d\mathbf{p} - \frac{5}{6m} \int \mathbf{p}^2 \varphi(\mathbf{p}) d\mathbf{p}$$

$$+ \frac{\hbar^2}{6m} \left( 3\rho_0 \frac{\partial \rho_c}{\partial \rho_0} + 2\theta \frac{\partial \rho_c}{\partial \theta} - \rho_c \right) \sum_{l,m,n} A^2 |c_{lmn}|^2 + \frac{\hbar^2 \rho_c \rho_0}{2m} \sum_{l,m,n} A^2 \frac{\partial |c_{lmn}|^2}{\partial \rho_0} + \frac{\hbar^2 \rho_c \theta}{3m} \sum_{l,m,n} A^2 \frac{\partial |c_{lmn}|^2}{\partial \theta}$$

$$- \frac{1}{2} \sum_{l,m,n} |a_{lmn}|^2 \left( \sigma_e(A) + \sigma(A) - \rho_0 \frac{\partial \sigma_e}{\partial \rho_0} + \frac{A}{3} \frac{\partial \sigma}{\partial A} - \theta \frac{\partial \sigma}{\partial \theta} - \frac{\theta A}{3} \frac{\partial^2 \sigma}{\partial \theta \partial A} \right)$$

$$+\frac{\rho_0}{2}\sum_{l,m,n}\frac{\partial|a_{lmn}|^2}{\partial\rho_0}\sigma_e(A)+\frac{\theta}{2}\sum_{l,m,n}\frac{\partial|a_{lmn}|^2}{\partial\theta}\left(\sigma(A)+\frac{A}{3}\frac{\partial\sigma}{\partial A}\right)=0. \tag{42}$$

To obtain an initial condition for this partial differential equation it is necessary first to solve Eq. (24) for $\tau(\theta,\rho_0)$ in the ordinary crystal. The values of this $\tau(\theta,\rho_0)$ taken on the line of the phase transition to the condensate crystal will serve as initial condition for (42).

The above thermodynamic functions depend not only on $\theta$ and $\rho_0$ but on $\rho_c$ as well, and therefore it needs an equation that would permit one to find $\rho_c$. In this question we can proceed from the fact that in a state of thermodynamic equilibrium the Helmholtz free energy $F$ is to be a minimum at given temperature $\theta$ and volume $V$ (or $\rho_0 = N/V$) [12]. Hence, we must minimize the free energy $F$ with respect to $\rho_c$. To find $F$ we observe that from (39) and (40) there results the relationship

$$2E - 3pV = VQ(\theta,\rho_0,\rho_c) \tag{43}$$

with

$$Q(\theta,\rho_0,\rho_c) = \sum_{l,m,n}|a_{lmn}|^2\left[\sigma_e(A) - \frac{3}{2}\sigma(A) - \frac{A}{2}\frac{\partial\sigma}{\partial A}\right]. \tag{44}$$

The energy $E$ and pressure $p$ are expressible in terms of the derivatives of $F$ by (25), which gives the following differential equation for the Helmholtz free energy $F$

$$2\theta\frac{\partial F}{\partial\theta}+3\rho_0\frac{\partial F}{\partial\rho_0}-2F=-VQ(\theta,\rho_0,\rho_c). \tag{45}$$

An equation of this type is solved in Appendix B of Ref. [11]. Analogously to Eq. (B10) of that appendix we get

$$F(\theta,\rho_0,\rho_c)=V\int_1^{\sqrt{\theta_0/\theta}}\frac{d\xi}{\xi^6}Q(\theta\xi^2,\rho_0\xi^3,\rho_c)+\frac{\theta}{\theta_0}F\left[\theta_0,\rho_0\left(\frac{\theta_0}{\theta}\right)^{3/2}\right], \tag{46}$$

wherein the last term contains the free energy at a fixed temperature $\theta_0$.

In case the temperature $\theta_0$ is so high that the condensate is absent, the last term in (46) will not contain $\rho_c$. In this instance, minimizing $F$ yields

$$\frac{\partial}{\partial\rho_c}\int_1^{\sqrt{\theta_0/\theta}}\frac{d\xi}{\xi^6}Q(\theta\xi^2,\rho_0\xi^3,\rho_c)=0. \tag{47}$$

It should be noted that in Refs. [7] and [13] an analogue of $Q$ does not contain $\theta$ and $\rho_0$, so that an equation of the type (47) can be written in the form of equations (3.15) of [7] and (3.15) of [13] (the argument $\rho$ in those equations should be deleted).

Equation (47) determines the dependence of the condensate density $\rho_c$ upon $\theta$ and $\rho_0$. It plays another role as well. If one places the quantities of this section in (47) and, after differentiating, sets $\rho_c = 0$, one will obtain the line of BEC in the $\rho_0$–$\theta$ phase plane mentioned when discussing Eq. (42) (cf. Ref. [7]).

## 5. Bifurcation method

In order to demonstrate how the general equations obtained in the previous sections can be applied for a concrete crystal, we shall consider one of the methods for solving the equations, namely, the bifurcation method (the idea of the method is presented in I). For the sake of simplicity we shall imply the simplest, from the mathematical point of view, case of a simple cubic (SC) lattice with space group $O_h^1$. This lattice was one of the lattices considered in II.

### 5.1. Ordinary quantum crystal

Application of the bifurcation method to an ordinary quantum crystal was studied in Ref. [6]. Seeing that Eq. (14) completely coincides with (3.4) of [6], the procedure of solving Eq. (14) is analogous with the one developed in [6]. The distinction lies in Eq. (15) compared with equation (3.5) of [6]. Integrals resulting from (15) are simpler than the ones following from (3.5) of [6] and can be expressed in terms of the error function.

Equations (7.1)–(7.3) and (7.6) of [6] remain valid for the present problem. The equation that determines the constant $B$ in terms of $\rho_0 = a_0$ follows from (15) at $l = m = n = 0$:

$$\rho_0 = \frac{B}{(2\pi)^4 i\hbar^3} \int_{(\infty)} d\mathbf{p} \int_C b_0(\mathbf{p},z) e^{-z/\tau} dz . \qquad (48)$$

The constant $B$ can be sought by successive approximations: $B = B_0 + B_1 + B_2 + \ldots$ . This expansion replaces (7.7) of Ref. [6]. Substituting (7.1) of [6] into (48) yields at once

$$B_0 = \rho_0 \hbar^3 \left(\frac{2\pi}{m\tau}\right)^{3/2} e^{\rho_0 \sigma_0/\tau}, \qquad B_1 = 0, \qquad (49)$$

where $\sigma_0 = \sigma(0)$.

Instead of (7.10) of Ref. [6] we introduce the integrals

$$I_n(\mathbf{A}_1,\ldots,\mathbf{A}_n) = \frac{1}{(2\pi)^4 i\hbar^3} \int_{(\infty)} d\mathbf{p} \int_C \frac{e^{-z/\tau} dz}{h_0(0) h_0(\mathbf{A}_1) \cdots h_0(\mathbf{A}_n)} \qquad (50)$$

with $h_0(\mathbf{A})$ defined in (7.2) of [6]. Properties of these integrals are considered in Appendix B.

In place of (7.11) of Ref. [6], the bifurcation point is determined by the equation

$$1 - B_0 \sigma(A) \, I_1(A) = 0. \tag{51}$$

Since $I_1 < 0$ according to (B3), the equation can be satisfied only if $\sigma(A) < 0$ as in Ref. [6]. Analogously to that reference, the magnitude of **A** at which (51) holds will be denoted by $a$, the value of 6 coefficients $a_{lmn}$ that are non-zero in the first approximation will be denoted by $\alpha$.

Numerical analysis shows that the quantity $B_0 I_1(a)$ is a minimum at $\tau = 0.05541 \hbar^2 a^2 / m$, the minimum being equal to $-5.1390 \rho_0 m / (\hbar a)^2$. Therefore, the condition of (51) cannot be satisfied if

$$\left| \sigma_{\min} \right| < 0.1946 \frac{\hbar^2 a^2}{\rho_0 m}, \tag{52}$$

where $\sigma_{\min}$ is the minimal value of $\sigma(k)$ between points $A$ and $B$ in Fig. 2 of I and $a = k_A$. Inequality (52) will take place when the interaction between the atoms is weak as in helium. In this event, formation of a crystal with space group $O_h^1$ under study is impossible and another space group should be analysed. It is interesting to note that the classical bifurcation condition (6.3) of [10] can always be fulfilled at low temperatures. From the above numerical results it follows also that the crystallization cannot occur in the range $0 \le \tau < 0.05541 \hbar^2 a^2 / m$: it occurs either at greater $\tau$ or never.

If $\sigma(a)$ is denoted by $\sigma_a$ as in II, terms of the second order in (48) give

$$B_2 = -\frac{6}{\rho_0} B_0^2 \alpha^2 \sigma_a^2 I_2(0, a). \tag{53}$$

The following equations will be written in the Kirkwood approximation (see II) upon assuming that $\sigma(\nu a) = 0$ when $\nu > 1$. Instead of (7.13) of Ref. [6], we have now

$$\alpha_2^{(1)} = 2 B_0 \sigma_a^2 I_2(\mathbf{a}_1, \mathbf{a}_2) \alpha^2, \qquad \alpha_2^{(2)} = B_0 \sigma_a^2 I_2(a, 2a) \, \alpha^2. \tag{54}$$

Similarly to (7.14) of [6], the value of $\alpha$ is given by

$$\alpha^2 = -\frac{1}{Y} \left[ 1 - B_0 \sigma_a \, I_1(a) \right], \tag{55}$$

where

$$Y = -B_0 \sigma_a^3 \left[ \frac{12}{\rho_0} \tau B_0 I_2^2(0, a) + I_3(0, a, a) + 2 I_3(0, \mathbf{a}, -\mathbf{a}) + 8 I_3(0, \mathbf{a}_1, \mathbf{a}_2) + 4 I_3(\mathbf{a}_1, \mathbf{a}_2, \mathbf{a}_1 + \mathbf{a}_2) \right].$$

We turn now to (24) that contains the function $\varphi(\mathbf{p})$ of (17). To calculate $\varphi(\mathbf{p})$ we substitute (7.1) and (7.3) of Ref. [6] into (17) restricting ourselves to terms of the second order. The occurring integrals can be treated analogously with Appendix B. Omitting details the result is

$$\int_{(\infty)} \frac{\mathbf{p}^2}{2m} \varphi(\mathbf{p}) d\mathbf{p} = \frac{3}{2} \tau \rho_0 \left[ 1 + \frac{3 \sigma_a \alpha^2}{\tau \rho_0} \right] - \frac{3 \rho_0 \sigma_a^2 \alpha^2}{2\tau} \left[ 1 + \frac{8 + \beta^2}{2\beta} i \sqrt{\pi} e^{-\beta^2/4} \mathrm{erf}\left( i \frac{\beta}{2} \right) \right], \tag{56}$$

where $\beta = \hbar a / \sqrt{2m\tau}$. Differentiating (56) with respect to $\theta$ and $\rho_0$ one obtains other terms that enters into (24), the coefficients $a_{lmn}$ in this approximation being given by (55). We shall no longer discuss Eq. (24) because special investigation is required to solve it together with the equation for the relevant fluid that should provide an initial condition for its solution.

### 5.2. Condensate crystal

In this case, $a_{lmn}$ should be replaced by $a_{lmn}^{(n)}$ in Eq. (15) on account of (31), and (35) should be substituted into (14). It is convenient to introduce $\xi_{lmn}$ instead of $a_{lmn}^{(c)}$ according to

$$a_{lmn}^{(c)} = \xi_{lmn} \, a_{lmn}^{(n)}. \tag{57}$$

Then the equations of the preceding subsection can be applied to the normal fraction provided $a_{lmn}$ is replaced by $a_{lmn}^{(n)}$ and $\rho_0$ by $\rho_n = \rho_0 - \rho_c$ while $\sigma(A)$ should be supplemented with the factor $(1 + \xi_{lmn})$. It will be noted that the combination $\rho_0\sigma_0$ remains unaltered. The method of the preceding subsection enables one to calculate the coefficients $a_{lmn}^{(n)}$ up to any desired order. The expressions thus obtained will contain the yet unknown quantities $\xi_{lmn}$.

We turn now to Eq. (33) that describes the condensate. As above, we exploit the Kirkwood approximation, in which case the potential $U(\mathbf{r})$ will be given by an equation analogous to (II.13):

$$U(\mathbf{r}) = \rho_0 \, \sigma_0 + 2\left(\alpha_1^{(c)} + \alpha_1^{(n)}\right)\sigma_a (\cos ax + \cos ay + \cos az), \tag{58}$$

in which $\alpha_1^{(c)}$ and $\alpha_1^{(n)}$ have the same meaning as $\alpha_1$ in (II.13). Simultaneously, we write down the first terms in the Fourier series for the condensate density:

$$\rho^{(c)}(\mathbf{r}) = \rho_c + 2\alpha_1^{(c)} (\cos ax + \cos ay + \cos az) + \ldots . \tag{59}$$

As long as we consider the case $\mathbf{p}_0 = 0$, Eq. (33) becomes

$$\frac{\hbar^2}{2m}\nabla^2 u(\mathbf{r}) + \left[\varepsilon_{(1)} - \rho_0\sigma_0 - 2\left(\alpha_1^{(c)} + \alpha_1^{(c)}\right)\sigma_a (\cos ax + \cos ay + \cos az)\right]u(\mathbf{r}) = 0. \tag{60}$$

This last equation is of the same form as (II.16), and its solution can be sought in the form

$$u(\mathbf{r}) = f_1(x) f_2(y) f_3(z). \tag{61}$$

Putting this into (60) we shall obtain an equation for $f_1(x)$ analogous with (II.18) and even the normalization of $f_1(x)$ will be the same (cf. (34) and (II.3)). Therefore, the required solution that corresponds to a minimum energy will be given by Eq. (II.23), namely,

$$f_1(x) = \sqrt{2}ce_0\left(\frac{ax}{2}, 4q\right), \tag{62}$$

but in the present instance

$$q = \frac{2m\sigma_a}{\hbar^2 a^2}\left(\alpha_1^{(c)} + \alpha_1^{(n)}\right). \tag{63}$$

Next, (61) with (62) should be placed in (32) and the result obtained should be expanded in a Fourier series. A comparison between the series and (59) will yield an equation for $\alpha_1^{(c)}$. To illustrate the procedure let us consider small $q$, in which case (62) gives, analogously to (II.24),

$$\rho^{(c)}(\mathbf{r}) = \rho_c\left[1 - \left(4q - 14q^3 + ...\right)(\cos ax + \cos ay + \cos az) + ...\right]. \tag{64}$$

Comparing this with (59) and substituting (63) yields

$$2\alpha_1^{(c)} = -\rho_c\left[\frac{8m\sigma_a}{\hbar^2 a^2}\left(\alpha_1^{(c)} + \alpha_1^{(n)}\right) - \frac{112m^3\sigma_a^3}{\hbar^6 a^6}\left(\alpha_1^{(c)} + \alpha_1^{(n)}\right)^3 + ...\right]. \tag{65}$$

One can look for a solution of the form

$$\alpha_1^{(c)} = \rho_c\left(\zeta_0 + \zeta_1\rho_c + \zeta_2\rho_c^2 + ...\right). \tag{66}$$

Substituting this into (65) gives

$$\zeta_0 = -\frac{4m\sigma_a}{\hbar^2 a^2}\alpha_1^{(n)}\left[1 - \frac{14m^2\sigma_a^2}{\hbar^4 a^4}\left(\alpha_1^{(n)}\right)^2 + ...\right], \quad \zeta_1 = -\frac{4m\sigma_a}{\hbar^2 a^2}\zeta_0\left[1 - \frac{42m^2\sigma_a^2}{\hbar^4 a^4}\left(\alpha_1^{(n)}\right)^2 + ...\right]. \tag{67}$$

Upon putting (66) into (57) we can find the quantities $\xi_{lmn}$. For example,

$$\xi_{\pm 100} = \xi_{0\pm 10} = \xi_{00\pm 1} = -\frac{4m\sigma_a\rho_c}{\hbar^2 a^2}\left[1 - \frac{14m^2\sigma_a^2}{\hbar^4 a^4}\left(\alpha_1^{(n)}\right)^2 + ...\right]\left[1 - \frac{4m\sigma_a}{\hbar^2 a^2}\rho_c + ...\right]. \tag{68}$$

Other quantities $\xi_{lmn}$ can be found by comparing further terms in (59) and (64). They will be expressed in terms of $a_{lmn}^{(n)}$ and $\rho_c$ like (68). Now one needs to revert to the equations discussed at the outset of this subsection and to put there the quantities $\xi_{lmn}$ obtained. This will permit one to find the coefficients $a_{lmn}^{(n)}$ in a final form, the coefficients $a_{lmn}^{(c)}$ being given by (66). The solution by the bifurcation method of the equations describing the condensate crystal will be finished, the expressions thus obtained containing $\rho_c$ as a parameter.

This solution should be placed in Eq. (42) that specifies the function $\tau(\theta,\rho_0,\rho_c)$. The first terms in the equation can be computed with the help of a formula analogous to (56). The equation for $\tau(\theta,\rho_0,\rho_c)$ should be solved together with the relevant equation for the ordinary crystal discussed at the end of the preceding subsection in order to have an initial condition for $\tau(\theta,\rho_0,\rho_c)$ at $\rho_c = 0$.

It remains to consider Eq. (47) that settles the equilibrium value of $\rho_c$. The function $Q(\theta,\rho_0,\rho_c)$ of (44) in the present Kirkwood approximation takes the form

$$Q(\theta,\rho_0,\rho_c) = 2\pi\rho_0^2 \int_0^\infty \left(2K + r\frac{dK}{dr}\right)g(r)r^2 dr + 6\left(\alpha_1^{(c)} + \alpha_1^{(n)}\right)^2\left[\sigma_e(a) - \frac{3}{2}\sigma(a) - \frac{a}{2}\frac{\partial\sigma}{\partial a}\right]. \quad (69)$$

The first term here has been transformed with use made of (13), (10) and (18). It is worthy of remark that, in the event of a fluid, the last summand is absent and the resulting $Q(\theta,\rho_0,\rho_c)$ when substituted into (46) yields an expression for the free energy that coincides with (B10) of Ref. [11]. In the fluid, the pair correlation function $g(r)$ must depend on $\rho_c$ otherwise the derivative of (47) will be nil identically. As to the crystal, the pair correlation function plays a secondary role as noted in Sec. 2 and we may even neglect the $\rho_c$-dependence of $g(r)$. In this approximation it is sufficient to consider solely the second summand of (69). We put the expressions for $\alpha_1^{(c)}$ and $\alpha_1^{(n)}$ obtained by the above method in the summand. The expressions will contain $\tau(\theta,\rho_0,\rho_c)$. Therefore, it needs to solve Eq. (42) for $\tau(\theta,\rho_0,\rho_c)$. After this we will be able to calculate the derivative of (47).

Closing the section it should be remarked that in the section we have only outlined the procedure of treating BEC in a crystal because to obtain concrete results it needs first to solve rather involved equations describing the ordinary quantum crystal. These last equations should be solved by a method other than the bifurcation method as long as the ordinary crystal is unstable in the vicinity of the bifurcation point at least in the classical case [10,9].

## 6. Absolute zero of temperature

In this section we revert to the case of zero temperature upon presuming that the condensate comprises all particles, the case being considered in detail in II. In the present paper we shall resort to a mathematical method that will enable us to arrive at new results concerning properties of a condensate crystal at zero temperature in addition to the ones obtained in II.

Below we shall imply Eq. (II.34), namely,

$$\frac{d^2 f_1}{d\xi^2} + 2ip\frac{df_1}{d\xi} + (c - 2q\cos\xi)f_1(\xi) = 0. \quad (70)$$

The equation describes the condensate in the case of a SC lattice (space group $O_h^1$) in the Kirkwood approximation (we conserve the premises and notation of II). In II, we solved the equation upon presuming that the parameter $q$ is small. If $p = 0$, the equation reduces to the well-known Mathieu equation. There exist two methods for solving the Mathieu equation if the parameter $q$ is large [14]. The first method yields solutions that have essential singularities and thereby are unfit for our purposes. We exploit the idea of the second method and apply it to (70).

First of all, we change the argument to $s = 2\chi\sin(\xi/2)$ where $\chi = (-q)^{1/4}$ (recall that $q < 0$ in our case). Next, we introduce a new function $u(s)$ according to $f_1 = u(s)\exp(-s^2/2)$. From (70) it follows the equation for $u(s)$

$$\left(4\chi^2 - s^2\right)\frac{d^2u}{ds^2} - \left(8s\chi^2 + s - 2s^3 - 8ip\sqrt{\chi^2 - \frac{s^2}{4}}\right)\frac{du}{ds}$$

$$+ \left(4c + 8\chi^4 - 4\chi^2 - 8ips\sqrt{\chi^2 - \frac{s^2}{4}} + 2s^2 - s^4\right)u = 0. \tag{71}$$

We imply the interval $-\pi \leq \xi \leq \pi$ where the radical in (71) is positive. The radical should be expanded in powers of $s$:

$$\sqrt{\chi^2 - \frac{s^2}{4}} = \chi - \frac{s^2}{8\chi} - \frac{s^4}{128\chi^3} - \frac{s^6}{1024\chi^5} - \ldots. \tag{72}$$

We seek a solution to (71) in the form

$$u = 1 + \frac{u_1(s)}{\chi} + \frac{u_2(s)}{\chi^2} + \frac{u_3(s)}{\chi^3} + \frac{u_4(s)}{\chi^4} + \ldots. \tag{73}$$

The function $u(s)$ is so normalized that no constant term is present in $u_i(s)$ (cf. II). The terms of (71) in $\chi^4$ and $\chi^2$ show that $c = -2\chi^4 + \chi^2$ in first approximations. Since the constant $c$ is to be real, it should be expanded as

$$c = -2\chi^4 + \chi^2 + \mu_1 + \frac{\mu_2}{\chi^2} + \frac{\mu_3}{\chi^4} + \frac{\mu_4}{\chi^6} + \ldots. \tag{74}$$

The terms of (71) linear in $\chi$ yield the equation

$$\frac{d^2u_1}{ds^2} - 2s\frac{du_1}{ds} - 2ips = 0. \tag{75}$$

As noted above, no constant term should be present in $u_i(s)$. Besides, the function $f_1 = u(s)\exp(-s^2/2)$ should be integrable, which is possible in particular if $u_i(s)$ are polynomials. As a result, $u_1 = -ips$. The terms of (71) in $\chi^0$ yield

$$4\frac{d^2u_2}{ds^2} - 8s\frac{du_2}{ds} + 8ip\frac{du_1}{ds} + 2s^2 - s^4 - 8ipsu_1 + 4\mu_1 = 0. \tag{76}$$

The constant terms specify $\mu_1$, and one gets

$$u_2 = -\frac{s^4}{32} + \frac{1-16p^2}{32}s^2, \qquad \mu_1 = -\frac{1}{16} - p^2. \tag{77}$$

In analogous way, one can calculate other terms in (73) and (74). For example,

$$u_3 = ip\left(\frac{s^5}{32} - \frac{7-16p^2}{96}s^3\right), \quad u_4 = \frac{s^8}{2048} - \left(\frac{5}{16} - p^2\right)\frac{s^6}{64} + \left(\frac{7}{256} - \frac{11p^2}{24} + \frac{p^4}{3}\right)\frac{s^4}{8} + \frac{s^2}{512}, \tag{78}$$

$$\mu_2 = -\frac{1}{256}, \quad \mu_3 = -\frac{3}{4096}. \tag{79}$$

The next step is to normalize $f_1$ according to (II.20) ($|f_1|^2$ should be taken instead of $f_1^2$). We write $f_1 = C_0 u \exp(-s^2/2)$ and transform the integrand in (II.20) to the variable $s$, so that

$$\frac{C_0^2}{2\pi\chi} \int_{-2\chi}^{2\chi} e^{-s^2} |u(s)|^2 \frac{ds}{\sqrt{1-\frac{s^2}{4\chi^2}}} = 1. \tag{80}$$

Herein the integrand diverges at the limits of integration. However, the contribution to the integral of regions near the limits of integration is exponentially small owing to the factor $\exp(-s^2)$ insofar as we imply that $\chi = (-q)^{1/4}$ is large. The singularity disappears once the radical is expanded like (72). Neglecting exponentially small terms one can now integrate from $-\infty$ to $+\infty$, in which case the occurring integrals are readily computed to yield

$$1/C_0^2 = \frac{1}{2\sqrt{\pi}\chi}\left(1 + \frac{3}{64\chi^2} + \frac{61}{8192\chi^4} + \ldots\right). \tag{81}$$

It remains now to deduce the equation for the coefficient $\alpha_1$ that figures in $q$ and thereby in $\chi$ in view of (II.21). The relevant procedure is analogous with the one of Sec. 4.3 of II. In the present instance, we should expand $|f_1(\xi)|^2$ in a Fourier series and find the coefficient $a_1$ of $\cos\xi$. Taking the relevant formula for $a_1$ and transforming the integrand similarly to (80) we get

$$a_1 = \frac{1}{\pi}\int_{-\pi}^{\pi} |f_1(\xi)|^2 \cos\xi d\xi = \frac{1}{\pi\chi}\int_{-2\chi}^{2\chi} |f_1(s)|^2 \left(1 - \frac{s^2}{2\chi^2}\right) \frac{ds}{\sqrt{1 - \frac{s^2}{4\chi^2}}}. \tag{82}$$

The first term (if one takes 1 in the parentheses) coincides with the left-hand side of (80) multiplied by 2. Proceeding in a like manner with the second term we arrive at

$$a_1 = 2 - \frac{1}{2\chi^2} + O\left(\frac{1}{\chi^6}\right). \tag{83}$$

Seeing that $|f_1|^2$ determines the crystal density $\rho(\mathbf{r})$ by (II.17) and (II.1), we compare the result with (II.14) and obtain, upon neglecting terms of order $1/(-q)^{3/2}$, the following equation for $\alpha_1$

$$\frac{\alpha_1}{\rho_0} = 1 - \frac{1}{4}\sqrt{\frac{-\hbar^2 a^2}{2m\alpha_1 \sigma_a}}. \tag{84}$$

The equation is valid when $|\sigma_a|$ is large. It is analogous to Eq. (II.92) for a FCC lattice.

This solution obtained enables one to deduce two important consequences. First, let us calculate the energy of the crystal given by (II.15). The quantity $\varepsilon_{(1)}$ of (II.33) which figures in (II.15) contains the term $\mathbf{p}_0^2/(2m)$ that is energetically disadvantageous when $\mathbf{p}_0 \neq 0$. If we calculate $c_1$ of (II.21) with use made of (74) (the expressions for $c_2$ and $c_3$ will be analogous), we shall see that the term $\mathbf{p}_0^2/(2m)$ disappears owing to $-p^2$ in $\mu_1$ of (77) (a similar phenomenon for the FCC lattice was noted in Sec. 4.3 of II). Even if $\mu_2$ and $\mu_3$ of (79) are taken into account, the crystal energy will not depend upon $\mathbf{p}_0$. Of course, the energy must depend on $\mathbf{p}_0$. This dependence, however, will be weak at least at small $p_0 = |\mathbf{p}_0|$ because of the factors $1/\chi^n$ in (74). It may turn out that the energy will be a minimum at $p_0 \neq 0$, in which case we shall have a superfluid crystal. Even if, in the Kirkwood approximation used, the minimum of the energy always corresponds to $p_0 = 0$, account taken of the quantities $\sigma(\nu a)$ with $\nu > 1$ may well change the situation. As a result, we can see that the existence of superfluid crystals (in the ground state, see II) is quite plausible.

The second consequence of the above solution consists in that the solution enables one to strictly calculate the momentum flow in the crystal. The momentum of the crystal is given by (I.27). For simplicity sake we shall assume that the vector $\mathbf{p}_0$ is directed along the $x$-axis. We substitute (II.17) and integrate over $y$ and $z$ with use made of the normalization of (II.3). As a result, the momentum along the $x$-axis is ($\rho_c = \rho_0$ in the present instance)

$$P_x = \rho_0 p_{0x} V - i\hbar \rho_0 V \frac{a}{2\pi} \int_{-\pi/a}^{\pi/a} f_1^*(x) \frac{df_1}{dx} dx. \tag{85}$$

We write $x = a\xi$ as in II, put $f_1 = C_0 u \exp(-s^2/2)$ as above, and transform the integral as in (80) and (82). The derivative $df_1/ds$ will give a term $s|u|^2$ which is an odd function of $s$ and disappears after integrating. As a result, we arrive at

$$P_x = \rho_0 p_{0x} V - i\hbar \rho_0 V C_0^2 \frac{a}{2\pi} \int_{-2\chi}^{2\chi} e^{-s^2} u^* \frac{du}{ds} ds. \tag{86}$$

As above, one can integrate from $-\infty$ to $+\infty$ upon neglecting exponentially small terms.

Calculating $u_i$ of (73) up to $u_7$ inclusive shows that there exists the relation

$$u^* \frac{du}{ds} = -ip \frac{|u|^2}{\chi\sqrt{1 - \frac{s^2}{4\chi^2}}} + F_{\text{odd}}(s), \tag{87}$$

where $F_{\text{odd}}(s)$ is an odd function of $s$. Substituting (87) into (86) with account taken of (80) yields $P_x = 0$ since $p = p_{0x}/(\hbar a)$. We could not prove that (87) holds in all orders although this is very probable, in which case Eq. (86) will always give $P_x = 0$. This does not signify however that the momentum of the crystal is zero identically. In the above calculations, we neglected exponentially small terms. As seen from (86) the terms are of the order $\exp(-4\chi^2) = \exp[-4(-q)^{1/2}]$. It is clear that the last factor determines the crystal density in interstices. The fact that the momentum flow is proportional to the density between adjacent crystal sites is understandable from the physical point of view.

## 7. Concluding remarks

In the present paper, mathematical methods were developed which enable one to study perfect crystals below the BEC point at arbitrary temperatures. We have also constructed thermodynamics describing the crystals in which a condensate is formed. The thermodynamics of an ordinary quantum crystal worked out in parallel with use made of the same approach provides a means of investigating the phase transition itself, namely, BEC in the crystals. It should be emphasized that BEC may occur in different crystals and the results of the paper open up new opportunities for detailed study of BEC and of its influence on properties of the crystals.

The analysis carried out with the help of the Landau theory of phase transitions shows that two situations are possible in the matter of the phase transition involving BEC. In the first situation, there are two different phase transitions. The first one is BEC where a condensate appears without superfluidity while the second one is a transition into a superfluid state (the second phase transition may not happen down to zero temperature). In the second situation, there is only one phase transition, namely, BEC where the system becomes simultaneously superfluid. The second situation is realized in liquid helium that becomes superfluid precisely at the $\lambda$-transition.

We have also demonstrated how the general equations obtained can be applied for a concrete crystal using the bifurcation method as an example. The bifurcation method, however, provides only limited possibilities for studying the equations describing the crystal. Therefore, it is desirable to develop other methods for handling the general equations of the paper, for which the method considered in Appendix A may be of utility.

In this paper, we have considered a mathematical method that enables one to find out extra properties of the condensate crystal at zero temperature in addition to the ones obtained in II. The energy of a superfluid crystal contains a term which, at first sight, renders the superfluid state energetically disadvantageous. In Sec. 6 it is shown that the term cancels out together with another term, so that the ground state of the condensate crystal may well be superfluid depending upon the interatomic interaction.

The next result obtained in Sec. 6 concerns the magnitude of the superflow. The superflow is proportional to the density between adjacent crystal sites which is quite understandable from the physical point of view. Ordinarily, the crystal density in the interstices is very slight, in which case the superflow will be difficult to observe experimentally. Only in solid helium where overlap of the wavefunctions of the adjacent sites is not small as compared with other crystals, is the superflow already observed (see I for relevant references). It should be remarked that, at high pressures when the distance between the atoms diminishes, the superflow should augment. At the same time, when the temperature is lowered nearing absolute zero, the magnitude of the spontaneous superflow may even decrease although the condensate density should increase. This is due to the fact that the crystal density in the interstices diminishes on lowering the temperature.

To obviate any misunderstanding lets us discuss now the concept of the condensate. Many authors are of the opinion that the condensate consists of particles having exactly zero momentum (see, e.g., Refs. [15-17]). This is true only for an ideal gas. It may be added that the particles of the ideal gas can move freely throughout the volume $V$ while $V \to \infty$ in the thermodynamic limit, so that zero momentum does not contradict the uncertainty principle. In the general case, the definition of the condensate may be formulated as follows. The condensate is a part of the system, the part comprising particles that are in an exceptional state with respect to other particles whose distributions over positions and momenta are described by usual laws of statistical mechanics. In particular, this definition is fully consistent with the situation in the ideal gas. The particles of the condensate in the system form a unified subsystem characterized by off-diagonal long range order (ODLRO) which is described by Eq. (I.25). At the same time it should be underlined that one cannot distinguish between the particles of the condensate and the ones of the normal fraction owing to the indistinguishability of identical particles. It may be added that the condensate can be superfluid or nonsuperfluid while only the condensate can be superfluid. When the condensate is nonsuperfluid in the thermodynamically equilibrium state of a crystal, it can become superfluid in a metastable (excited) state (see II).

The distribution over momenta in the condensate of a crystal is found out in Ref. [18]. It consists of δ-peaks with dissimilar multipliers that are arranged periodically in momentum space, the periodicity being characterized by the basic reciprocal-lattice vectors $\mathbf{a}_1$, $\mathbf{a}_2$ and $\mathbf{a}_3$ multiplied by $\hbar$. The peaks are displaced from the origin by the vector $\mathbf{p}_0$. If the condensate crystal is not superfluid ($\mathbf{p}_0 = 0$), the periodic structure in momentum space remains.

To determine the real concentration of the condensate experimentally is not a simple matter. If, for example, one measures the dynamic structure factor of liquid helium as in Ref. [16], account must be taken of the remark at the end of Sec. 3 in Ref. [18]. Measurements of non-classical rotational inertia of a crystal (for references see I) do not yield the full condensate concentration either since the magnitude of the superflow in the crystal should be small even if the condensate comprises all particles (see above). Besides, the condensate crystal may turn out nonsuperfluid.

**Appendix A. Alternative method for an ordinary crystal**

Equation (3) can be treated with a method analogous to the one employed in Sec. 6 of Ref. [6]. In that section, however, the occurring functions were expressed in terms of Fourier integrals which should be handled with caution in case periodic functions are involved. In this appendix we shall utilize other expansions.

As in Ref. [6], instead of $v(\mathbf{r},\mathbf{p},z)$ we introduce a function $w(\mathbf{r},\mathbf{p},z)$ according to

$$v(\mathbf{r},\mathbf{p},z) = w(\mathbf{r},\mathbf{p},z)\exp(-i\mathbf{pr}/\hbar). \tag{A1}$$

From (3) the equation for $w(\mathbf{r},\mathbf{p},z)$ follows:

$$\frac{\hbar^2}{2m}\nabla^2 w + [z - U(\mathbf{r})]w = \exp\left(\frac{i\mathbf{pr}}{\hbar}\right). \tag{A2}$$

The homogeneous equation complementary to (A2) is

$$\frac{\hbar^2}{2m}\nabla^2 \psi_\nu(\mathbf{r}) + [\varepsilon_\nu - U(\mathbf{r})]\psi_\nu(\mathbf{r}) = 0. \tag{A3}$$

In distinction to [6], we shall imply solutions of this equation of the type

$$\psi_\nu(\mathbf{r}) = \widetilde{\psi}_\nu(\mathbf{r})\exp\left(\frac{i\mathbf{pr}}{\hbar}\right), \tag{A4}$$

in which the functions $\widetilde{\psi}_\nu(\mathbf{r})$ form a complete set of periodic orthonormal functions having the same periods as $U(\mathbf{r})$. It will be noted that the functions $\widetilde{\psi}_\nu(\mathbf{r})$ depend on $\mathbf{p}$ as well owing to (A4), and sometimes we shall write $\widetilde{\psi}_\nu(\mathbf{r},\mathbf{p})$ to underline this (and we shall write also $\varepsilon_\nu(\mathbf{p})$). We look for $w(\mathbf{r},\mathbf{p},z)$ in the form

$$w(\mathbf{r},\mathbf{p},z) = \sum_\nu \gamma_\nu(\mathbf{p},z)\psi_\nu(\mathbf{r}). \tag{A5}$$

Substituting (A5) into (A2) with account taken of (A3)–(A4) and of the orthogonality relations for $\widetilde{\psi}_\nu(\mathbf{r})$ leads to

$$\gamma_\nu(\mathbf{p},z) = \frac{1}{z-\varepsilon_\nu}\int_{v_0}\widetilde{\psi}_\nu^*(\mathbf{r})d\mathbf{r}. \tag{A6}$$

Thus, by (A1) the function $v(\mathbf{r},\mathbf{p},z)$ that figures in (2) can be represented as

$$v(\mathbf{r},\mathbf{p},z) = \sum_\nu \frac{\widetilde{\psi}_\nu(\mathbf{r},\mathbf{p})}{z - \varepsilon_\nu(\mathbf{p})} \int_{v_0} \widetilde{\psi}_\nu^*(\mathbf{r}',\mathbf{p}) d\mathbf{r}'. \qquad (A7)$$

The last integral is nonzero only if $\widetilde{\psi}_\nu(\mathbf{r})$ contains a constant term as the integral is proportional to the relevant Fourier coefficient (see, e.g., (II.89) with $\mathbf{A} = 0$). It is not difficult to demonstrate that only one of the functions $\widetilde{\psi}_\nu(\mathbf{r})$ has the constant term. To this end we tend $U(\mathbf{r})$ in (A3) to a constant $U_0$. In this limit, upon putting $\widetilde{\psi}_\nu(\mathbf{r}) =$ constant in (A4) and substituting this into (A3) we shall obtain only one eigenvalue

$$\varepsilon_\nu = U_0 + \frac{\mathbf{p}^2}{2m}, \qquad (A8)$$

and thereby only one eigenfunction (other eigenfunctions $\widetilde{\psi}_\nu(\mathbf{r})$ will be proportional to $\exp(i\mathbf{A}\mathbf{r})$ with $\mathbf{A} \neq 0$). If we denote the function $\widetilde{\psi}_\nu(\mathbf{r})$ with the constant term as $\widetilde{\psi}_0(\mathbf{r})$ and the relevant eigenvalue as $\varepsilon_0$, from (A7) we shall get

$$v(\mathbf{r},\mathbf{p},z) = v_0 \psi_0(\mathbf{p}) \frac{\widetilde{\psi}_0(\mathbf{r},\mathbf{p})}{z - \varepsilon_0(\mathbf{p})}, \qquad (A9)$$

where $\psi_0(\mathbf{p})$ is the constant term of $\widetilde{\psi}_0(\mathbf{r},\mathbf{p})$. Equation (A9) is much simpler than (6.7) of Ref. [6] which contains an infinite sum.

The function $b_0(\mathbf{p},z)$ that figures in (17) is a constant term in the series of (11) for $v(\mathbf{r},\mathbf{p},z)$. From (A9) we obtain at once

$$b_0(\mathbf{p},z) = \frac{v_0 \psi_0^2(\mathbf{p})}{z - \varepsilon_0(\mathbf{p})}. \qquad (A10)$$

Now we can readily integrate over $z$ in (17) with the help of the residue theorem:

$$\varphi(\mathbf{p}) = \frac{B}{(2\pi\hbar)^3} v_0 \psi_0^2(\mathbf{p}) e^{-\frac{\varepsilon_0(\mathbf{p})}{\tau}}. \qquad (A11)$$

Hence, the energy of (16), the stress tensor of (19) and the first terms in Eq. (24) for $\tau(\theta,\rho_0)$ are expressible in terms of $\psi_0(\mathbf{p})$ and $\varepsilon_0(\mathbf{p})$.

To avoid confusion, the following should be stressed. The functions $\psi_0(\mathbf{p})$ and $\varepsilon_0(\mathbf{p})$ for a crystal can be considered to be periodic functions in the reciprocal space [19]. This being so, integrals over $\mathbf{p}$ containing $\varphi(\mathbf{p})$ (for example, the integral in (16)) will not exist. In the present method, however, these functions are not periodic. In the limiting case of a uniform medium (a fluid) we should arrive at (A8). Therefore, the above function $\varepsilon_0(\mathbf{p})$ must be of the form presented in Fig. 12a of [19].

**Appendix B. Integrals for the bifurcation method**

The integrals $I_n(\mathbf{A}_1,\ldots,\mathbf{A}_n)$ of (50) have the same properties as in (B.1) of [6]. The integration over $z$ can be readily performed with the aid of residue theory. Thus, it remains to integrate over $\mathbf{p}$. If the vectors $\mathbf{A}_1,\ldots,\mathbf{A}_n$ have a common direction, the $p_x$-axis can be oriented in the same direction. Then the integrations over $p_y$ and $p_z$ are carried out at once. The integral over $p_x$ can be calculated with the help of the formula (see Sec. 7.1.4 of Ref. [20])

$$\int_{-\infty}^{\infty} \frac{e^{-t^2} dt}{t \mp z} = i\pi e^{-z^2} \mp 2\sqrt{\pi} e^{-z^2} \int_0^z e^{t^2} dt = i\pi e^{-z^2} \pm i\pi \operatorname{erf}(iz), \tag{B1}$$

if the path of integration passes below the singularity of the integrand. Other necessary formulae can be obtained by differentiating (B1) with respect to $z$.

When considering double integrals it is preferable to utilize the principal value of the integral

$$P \int_{-\infty}^{\infty} \frac{e^{-t^2} dt}{t \mp z} = \pm i\pi \operatorname{erf}(iz), \tag{B2}$$

which follows from (B1).

Omitting details of the intervening calculations sometimes rather involved we present the results in which $\beta = \hbar a / \sqrt{2m\tau}$.

$$I_1(a) = \frac{4\pi^2 m^2 \tau i}{(2\pi\hbar)^3 \hbar a} \exp\left(-\frac{\rho_0 \sigma_0}{\tau} - \frac{\beta^2}{4}\right) \operatorname{erf}\left(i\frac{\beta}{2}\right) = -\frac{8\pi^{3/2} m^2 \tau}{(2\pi\hbar)^3 \hbar a} \exp\left(-\frac{\rho_0 \sigma_0}{\tau} - \frac{\beta^2}{4}\right) \int_0^{\beta/2} e^{t^2} dt. \tag{B3}$$

$$I_2(0,a) = -\frac{1}{2\tau} I_1(a). \tag{B4}$$

$$I_2(\mathbf{a}_1, \mathbf{a}_2) = -\frac{2\pi^2 m^2 \sqrt{2\pi m \tau}}{(2\pi\hbar)^3 \hbar^2 a^2} \exp\left(-\frac{\rho_0 \sigma_0}{\tau} - \frac{\beta^2}{2}\right) \operatorname{erf}^2\left(i\frac{\beta}{2}\right). \tag{B5}$$

$$I_2(a,2a) = I_2(a,-a) = \frac{4\pi^2 m^3 \tau i}{(2\pi\hbar)^3 \hbar^3 a^3} \exp\left(-\frac{\rho_0 \sigma_0}{\tau}\right)\left[\exp(-\beta^2) \operatorname{erf}(i\beta) - 2\exp\left(-\frac{\beta^2}{4}\right) \operatorname{erf}\left(i\frac{\beta}{2}\right)\right]. \tag{B6}$$

$$I_3(0,a,a) = \frac{2\pi^{3/2} m^3}{(2\pi\hbar)^3 \hbar^3 a^3} \exp\left(-\frac{\rho_0 \sigma_0}{\tau}\right)\left[2\beta + i\sqrt{\pi}(2+\beta^2) \exp\left(-\frac{\beta^2}{4}\right) \operatorname{erf}\left(i\frac{\beta}{2}\right)\right]. \tag{B7}$$

$$I_3(0,\mathbf{a},-\mathbf{a}) = \frac{8\pi^2 m^4 \tau i}{(2\pi\hbar)^3 \hbar^5 a^5} \exp\left(-\frac{\rho_0 \sigma_0}{\tau}\right)\left[\exp(-\beta^2) \operatorname{erf}(i\beta) - (2-\beta^2) \exp\left(-\frac{\beta^2}{4}\right) \operatorname{erf}\left(i\frac{\beta}{2}\right)\right]. \tag{B8}$$

$$I_3(0,\mathbf{a}_1,\mathbf{a}_2) = \frac{8\pi^2 m^3 i}{(2\pi\hbar)^3 \hbar^3 a^3} \exp\left(-\frac{\rho_0 \sigma_0}{\tau}\right)\left[\exp\left(-\frac{\beta^2}{4}\right) \operatorname{erf}\left(i\frac{\beta}{2}\right) - \frac{1}{\sqrt{2}} \exp\left(-\frac{\beta^2}{2}\right) \operatorname{erf}\left(i\frac{\beta}{\sqrt{2}}\right)\right]. \tag{B9}$$

$$I_3(\mathbf{a}_1,\mathbf{a}_2,\mathbf{a}_1+\mathbf{a}_2) = \frac{2\sqrt{2}(\pi m)^{5/2}}{(2\pi\hbar)^3 \hbar^2 a^2 \sqrt{\tau}} \exp\left(-\frac{\rho_0\sigma_0}{\tau}-\frac{\beta^2}{2}\right) \mathrm{erf}^2\left(i\frac{\beta}{2}\right) - 2I_3(0,\mathbf{a}_1,\mathbf{a}_2). \qquad (B10)$$

When $\tau \to \infty$ ($\beta \to 0$), the limiting value of the integrals is given by

$$I_n(\mathbf{A}_1,\ldots,\mathbf{A}_n) \to \frac{(-1)^n}{n!\hbar^3\tau^n}\left(\frac{m\tau}{2\pi}\right)^{3/2} \exp\left(-\frac{\rho_0\sigma_0}{\tau}\right). \qquad (B11)$$

If $\tau \to 0$ ($\beta \to \infty$), upon representing the error function erf($ix$) as in (B3) the limiting behaviour of the integrals can be found out with the aid of the following asymptotic expansion as $x \to \infty$

$$e^{-x^2}\int_0^x e^{t^2}dt \sim \frac{1}{2x}\left[1+\sum_{k=1}^{\infty}\frac{1\cdot 3\ldots(2k-1)}{(2x^2)^k}\right]. \qquad (B12)$$

**References**

1. V.A. Golovko, Eur. Phys. J. B **71**, 85 (2009).
2. V.A. Golovko, Eur. Phys. J. B **74**, 345 (2010).
3. V.A. Golovko, Physica A **230**, 658 (1996).
4. V.A. Golovko, in *Progress in Statistical Mechanics Research*, edited by J. S. Moreno (Nova Science Publ., Hauppauge, NY, 2008), Chap. 2.
5. N. Prokof'ev, Adv. Phys. **56,** 381 (2007).
6. V.A. Golovko, Physica A **310**, 39 (2002).
7. V.A. Golovko, Physica A **246**, 275 (1997), see also the full version of this paper in [arXiv: 0902.4134].
8. V.A. Golovko, Physica A **374**, 15 (2007).
9. V.A. Golovko, Physica A **341**, 340 (2004).
10. V.A. Golovko, Physica A **300**, 195 (2001).
11. V.A. Golovko, J. Math. Phys. **44**, 2621 (2003).
12. L. D. Landau and E. M. Lifshitz, *Statistical Physics*, Part I (Pergamon, Oxford, 1980).
13. V.A. Golovko, J. Math. Phys. **45**, 1571 (2004).
14. D. Frenkel and R. Portugal, J. Phys. A: Math. Gen. **34**, 3541 (2001).
15. D.M. Ceperley, Rev. Mod. Phys. **67**, 279 (1995).
16. H.R. Glyde, R.T. Azuah and W.G. Stirling, Phys. Rev. B **62**, 14337 (2000).
17. S. Moroni and M. Boninsegni, J. Low Temp. Phys. **136**, 129 (2004).
18. V.A. Golovko, Momentum representation for equilibrium reduced density matrices, arXiv: 1105.3563 [quant-ph].
19. E.M. Lifshitz and L.P. Pitaevskii, *Statistical Physics*, Part 2 (Pergamon, Oxford, 1980), § 55.
20. M. Abramovitz and I.A. Stegun (Eds.), *Handbook of Mathematical Functions* (USGPO, Washington, 1964).